\documentclass[twocolumn,amsmath,floats,showpacs,nofootinbib]{revtex4}
\usepackage[usenames]{color}
\usepackage{graphicx}
\usepackage{epstopdf}
\usepackage{textcomp}
\usepackage{hyperref}
\usepackage{multirow}
\usepackage{tikz}
\usepackage{rotating}
\hypersetup{colorlinks=true}

\begin{document}

\title{Superconducting energy gap in $\rm Ba_{1-x}K_xBiO_3$: Temperature dependence}

\author{F. Szab$\rm \acute{o}^a$, P. Samuely$^a$, N.L. Bobrov$^b$, J. Marcus$^c$, C. Escribe-Filippini$^c$, and M. Affronte$^c$}
\affiliation{$^a$Institute of Experimental Physics, Slovak Academy of Sciences, CS-04353 Ko$\check{s}$ice, Slovakia\\
$^b$Institute for Low Temperature Physics and Engineering, Ukrainian Academy of Sciences, Kharkov, Ukraina\\
$^c$Laboratoire d'Etudes des Propri$\acute{e}t\acute{e}$s Electroniques des Solides CNRS, BP 166, F-38042 Grenoble Cedex 9, France\\
Email address: bobrov@ilt.kharkov.ua}

\published {\href{https://doi.org/10.1016/0921-4534(94)92158-X}{Physica C}, \textbf{235-240}, 1873 (1994)}
\date{\today}

\begin{abstract}The superconducting energy gap of $\rm Ba_{1-x}K_xBiO_3$ has been measured by tunneling. Despite the fact that the sample was macroscopically single phase with very sharp superconducting transition $T_c$ at 32~$K$, some of the measured tunnel junctions made by point contacts between silver tip and single crystal of $\rm Ba_{1-x}K_xBiO_3$ had lower transition at 20~$K$. Local variation of the potassium concentration as well as oxygen deficiency in $\rm Ba_{1-x}K_xBiO_3$ at the place where the point contact is made can account for the change of $T_c$. The conductance curves of the tunnel junctions reveal the BCS behavior with a small broadening of the superconducting-gap structure. A value of the energy gap scales with $T_c$. The reduced gap amounts to $2\Delta/kT_c = 4\div 4.3$ indicating a medium coupling strength. Temperature dependence of the energy gap follows the BCS prediction.
\end{abstract}

\maketitle

Bismuthate superconductors in contrast to the cuprates with a quasi twodimensional lattice, are fully 3-dimensional with cubic symmetry and diamagnetism in the normal state. Their superconducting properties seem to be understood within the classical theory. Tunneling studies on $\rm Ba_{1-x}K_xBiO_3$ have shown a full superconducting energy gap $\Delta$ with the reduced value $2\Delta/kT_c$ ranging from the weak coupling limit \cite{1,2} to the medium coupling \cite{3}. It is generally accepted that the electron-phonon interaction plays a role in the superconductivity here \cite{1,3}. There is on the other hand some similarity with the cuprates. Both perovskites are near the metal-insulator transition triggered by doping. Namely, for $\rm Ba_{1-x}K_xBiO_3$ the system becomes metallic (supeconducting) at $x\sim 0.35$. The highest transition temperature $T_c=32\ K$ is achieved near the metal-insulator transition and then it is decreased down to 20~$K$ for $x=0.5$, the solubility limit. Asymmetric linear background of the tunneling conductance may indicate strong electronic correlations in the normal state.

The $\rm Ba_{1-x}K_xBiO_3$ crystals used in this experiment were grown by electrochemical method \cite{4}. They are characterized by the high and sharp superconducting transition at $T_c=32\ K$. They are macroscopically single phase. The point-contact technique has been used to make the tunnel junctions with a silver single crystal as a tip.
\begin{figure}[h!]
\includegraphics[width=8.5cm,angle=0]{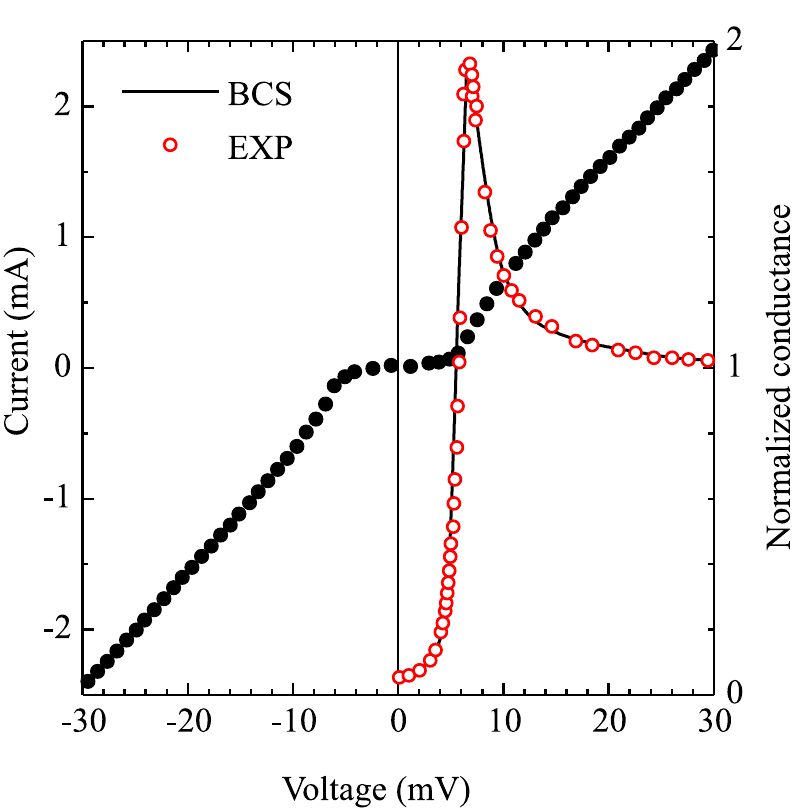}
\caption[]{Tunneling conductance at 4.2~$K$ of the $\rm Ba_{1-x}K_xBiO_3-Ag$ junction with $T_c=32\ K$ and the fit by the BCS density of states.}
\label{Fig1}
\end{figure}

\begin{figure}[]
\includegraphics[width=8.5cm,angle=0]{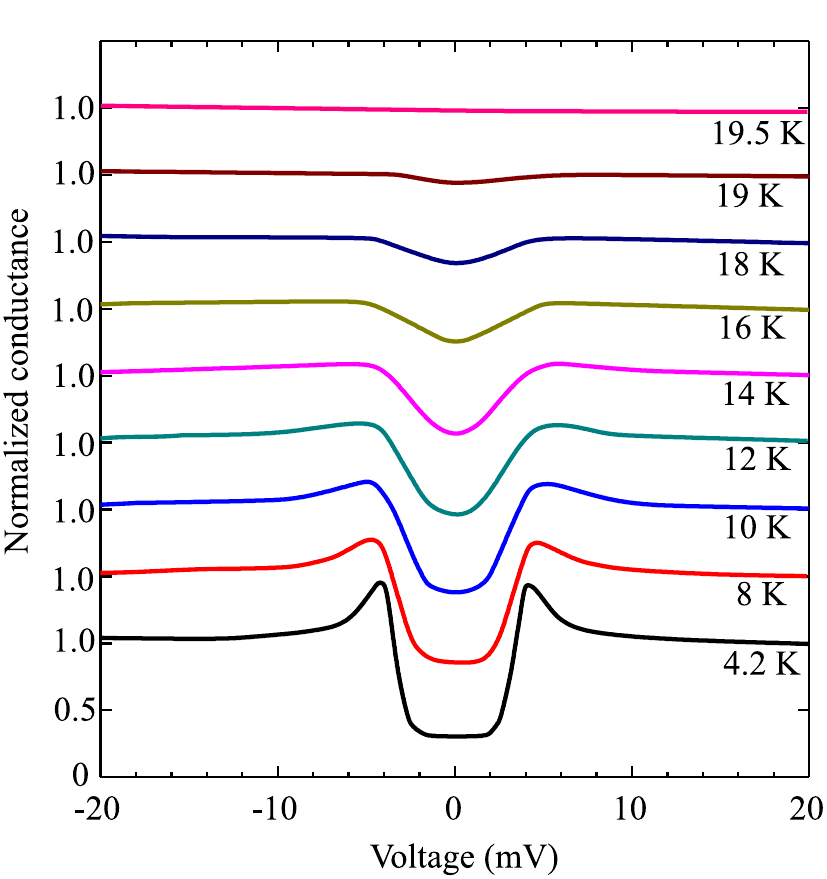}
\caption[]{Temperature dependence of the spectrum for the tunnel junction with $T_c=20\ K$.}
\label{Fig2}
\end{figure}

\begin{figure}[]
\includegraphics[width=8.5cm,angle=0]{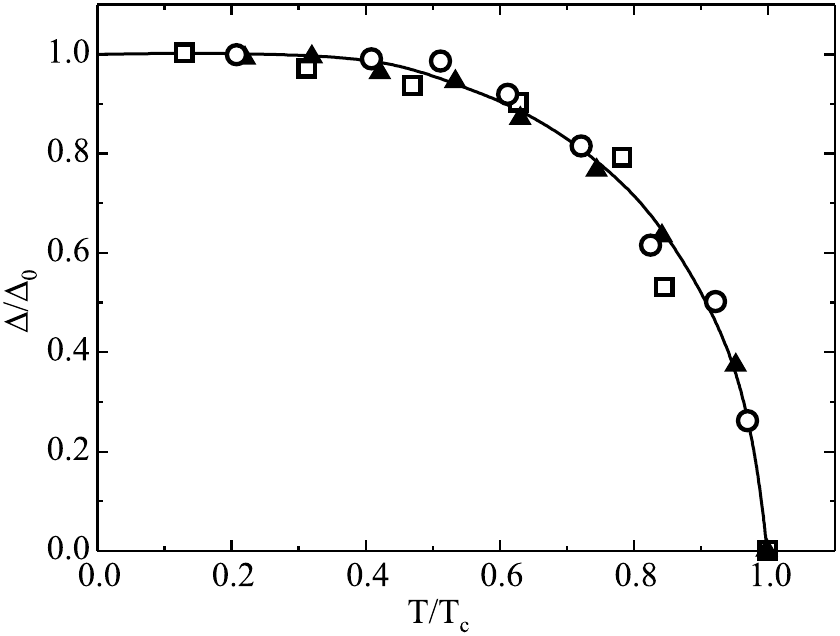}
\caption[]{Temperature dependence of the super-conducting energy gap. Dashed line - BCS curve.}
\label{Fig3}
\end{figure}

Figure \ref{Fig1} shows a typical tunneling conductance trace with the single gap structure, which could
be fitted by the BCS density of states $N(E)=E/\sqrt{{{E}^{2}}-{{\Delta }^{2}}}$ at 4.2~$K$, taking into account a smearing factor $\Gamma$ by replacement $E$ by $E' + i\Gamma$ as the only extra parameter (Dynes formula). Actually, the superconducting energy gap $\Delta$ equals to 6~$meV$ and very small smearing factor $\Gamma = 0.35\ meV$, $T_c$ of the tunnel junction was 32~$K$.

We measured also the temperature dependence of the tunneling effect. In few cases we found the transition temperature of the tunnel junction different from the bulk $T_c$. As shown in Fig.\ref{Fig2}, the transition $T_c$ was achieved at about 20~$K$. Lower local $T_c$ can be caused by a presence of microphases of different stoichiometry, e.g. by variation in the concentration of potassium and/or the oxygen deficiency. Local deviations in stoichiometry seem to be a general problem of the bismuthates. It is worth noticing that our sample does not show multiphase character in acsusceptibility and it has a high metallic conductance above $T_c$ \cite{4}. We fitted the experimental data by the Dynes formula with resulting values: $\Delta_0=3.5\ meV$, $\Gamma=0.5\ meV$.

In Fig.\ref{Fig3} the temperature dependence of the superconducting energy gap obtained from the data of three different junctions is displayed in the reduced coordinates to account for different $T_c$, resp. $\Delta_0$. In all three cases the data follow the BCS prediction.

The reduced superconducting energy gap $2\Delta/kT_c$ amounts to $4\div 4.3$ for all junctions. Hence the gap scales with the $T_c$ in $\rm Ba_{1-x}K_xBiO_3$. Presence of microdomains of different phases observed by our point-contact method may affect several physical properties measured in the system.

This work was partially supported by the Commision of the European Communities Contract No.CIPA-CT93-0183.

\end{document}